\pdfoutput=1
\documentclass{article}
\usepackage{PRIMEarxiv}

\usepackage[utf8]{inputenc}
\usepackage[T1]{fontenc}

\usepackage{hyperref}
\usepackage{url}
\usepackage{booktabs}
\usepackage{tabularx}
\usepackage{array}
\usepackage{amsfonts}
\usepackage{amsmath}
\usepackage{microtype}
\usepackage{graphicx}
\usepackage[numbers,sort&compress]{natbib}
\usepackage{float}
\usepackage{xcolor}
\usepackage{subcaption}
\usepackage{tikz}
\usetikzlibrary{arrows.meta,positioning,calc,fit,backgrounds}
\usepackage{fancyhdr}
\usepackage{titlesec}
\titlespacing*{\section}{0pt}{1.2ex}{0.8ex}

\title{The Context Flywheel: Growing an Enterprise Security Brain\\from Its Own Investigations}

\author{
  \textbf{Boris Plotnikov}, \textbf{Ben Benhemo}, \textbf{Gal Engelberg}, \textbf{Leon Goldberg} \\
  Sola Security \\
  \texttt{Corresponding author: gal.e@sola.security}
}
\date{}

\hypersetup{
  pdftitle={The Context Flywheel: Growing an Enterprise Security Brain from Its Own Investigations},
  pdfauthor={Boris Plotnikov, Ben Benhemo, Gal Engelberg, Leon Goldberg},
  pdfkeywords={enterprise security, security knowledge graph, investigation agents, query mining, human-in-the-loop, LLM-as-a-judge},
  colorlinks=true, linkcolor=blue!50!black, citecolor=blue!50!black, urlcolor=blue!50!black,
}

\begin{document}
\maketitle

\begin{abstract}
Security agents increasingly investigate an organization through an enterprise security brain, a \emph{context layer}: a shared model of its assets and their relationships, built in advance so agents do not reconstruct them from raw product data for each question. The brain can answer only through the relationships it already models. When an investigation needs one it lacks, such as which deployment created a vulnerable function, the agent works it out with its own query, uses it once and discards it, so the next investigation repeats the work. We call this the \emph{modeling gap}. Closing it raises three challenges: \emph{effectiveness}, to what extent a relationship learned in one organization can be used in the model all organizations share; \emph{precision}, how many of the agents' proposals are correct when one wrong relationship misleads every later investigation; and \emph{scale}, admitting them without expert review growing with every investigation. We present the \emph{context flywheel}, a system that increases the brain's knowledge with use. It identifies the relationship patterns agents work out, removes malformed and already-modeled candidates automatically, and sends the rest to an expert; each accepted pattern then serves every later investigation in every organization. After each review, the expert's recurring reasons for rejection become rules that remove the same mistakes automatically and guide a language-model judge, so later reviews start from fewer candidates. We evaluated the flywheel on 656 candidate patterns proposed in 21 production environments. The expert accepted 42\%. The accepted patterns brought 49 previously unlinked asset types into the brain's type-level graph, raising its global efficiency by 13\%. Rules written from this review would have removed a quarter of the queue while losing 2.5\% of the accepted patterns.
\end{abstract}

\keywords{Enterprise security \and Security knowledge graphs \and Investigation agents \and Query mining \and Human-in-the-loop \and LLM-as-a-judge}

\section{Introduction}

Language-model agents now investigate enterprise security, querying an organization's data to answer questions about its environment and to decide what a security finding means \citep{goldberg2026coding,goldberg2026findings}. Yet the data an investigation needs is spread across products. An organization runs workloads in one or more cloud providers, manages identities in an identity provider, keeps code and deployment pipelines on a code platform, and receives findings from several scanners, each describing its own assets in its own schema. Most investigative questions cross these boundaries: which cloud roles a code repository can assume, which identity provider authenticates a load balancer, which deployment created a vulnerable function. An agent working from raw product data reconstructs each such connection when the question arrives, and pays that cost again on the next investigation \citep{goldberg2026coding}.

A \emph{context layer} removes that reconstruction. It is a model of the environment, prepared before questions arrive, that agents read instead of raw product data; attack graphs \citep{sheyner2002}, logical attack models \citep{ou2005mulval} and security knowledge graphs \citep{sikos2023} are earlier models of this kind. The \emph{enterprise security brain} is a context layer built for agents \citep{engelberg2026asset,yavin2026context}. It holds the organization's assets from every connected product as nodes of one graph, connected according to \emph{link types}: relationship patterns between two asset types, such as ``role can assume role'' or ``stack deploys function'', where a stack is the set of cloud resources created together from one template. A link type is a rule for connecting the two types, for instance by matching the stack identifier written into a function's tags. The brain's model is thus defined at the type level, over asset types and link types, and each organization's data manifests it at the instance level, as that organization's own graph of assets and link instances; a link type modeled once connects the matching assets in every organization whose data holds both types. Security knowledge, such as which controls an asset fails, is evaluated over the same graph. On cloud investigations about whole populations of assets, a brain of this kind covered more of the answer than general coding agents working from raw cloud APIs, at a fraction of their cost \citep{goldberg2026coding}.

The brain helps only where it holds the link types a question needs. On most of the tasks where it did not lead the coding agents, the answer depended on relations between assets that it had not modeled \citep{goldberg2026coding}. Its link types are added in advance by the brain's builders, from each vendor's schema, from identifiers that products share, and from relations that a language model proposes and that are checked against data \citep{engelberg2026asset}. Builders add a link type before any question needs it, so some link types are still missing when a question arrives.

The agents that read the brain find these missing link types themselves. An agent that judges a finding on one resource often needs data that another product holds about that resource: in production, a substantial share of such verdicts rested on another product's data that the brain does not link to the resource \citep{goldberg2026findings}. The agent then writes its own query to connect the two, uses it for that one investigation, and discards it, and the next investigation that needs the same connection writes it again. We call the distance between the link types the brain holds and the link types its agents work out the \emph{modeling gap}.

Closing the modeling gap from the side of the agents poses three challenges. The first is \emph{effectiveness}: a relationship pattern is found while settling one investigation in one organization, and the open question is to what extent it can be used in the model that all organizations share. The second is \emph{precision}: agents also propose patterns that connect the wrong assets, and a wrong link type in a shared model is read by every later investigation, while a missed link type can be proposed again. Admission must therefore favor precision, and which properties of a proposal signal that it is correct has to be measured. The third is \emph{scale}: expert review supplies that precision, and its cost grows with every investigation that proposes candidates.

We present the \emph{context flywheel}, a system that increases the knowledge of a context layer with use (Figure~\ref{fig:flywheel}). The flywheel is an automatic process with a human in the loop. Each relationship an agent works out during an investigation becomes a candidate link type, and automatic checks discard candidates that are malformed, specific to one organization, empty or already modeled. The one human step is admission: a security \emph{expert} accepts, edits or rejects each remaining candidate before it enters the shared model. Each accepted link type then serves every later investigation in every organization, and those investigations surface the next gaps. What the flywheel shares is a pattern at the type level, never data at the instance level. An accepted link type is a rule over asset types, such as matching a function to the stack named in its tags, and it carries no asset, value or record of the organization where it was found. Each organization evaluates the rule on its own data, so the knowledge travels across organizations while each organization's data stays its own.

The expert's time is the scarce resource, so the flywheel also learns from each review. Many rejections repeat the same mistake: the most common rejected candidate connects two asset types only because they sit in the same account. After a batch, the expert writes such recurring reasons down as short rules, and the flywheel applies them before the next review. A rule that can be read from the query becomes an automatic check that removes the candidate before the expert sees it; a language-model judge grades the remaining candidates against the rules; and the proposing agents receive the rules as instructions, so they propose fewer such candidates. The expert is then left with the candidates the rules cannot settle. We call this the \emph{inner loop}.

We evaluate the flywheel on a sample of one production review batch: candidate link types that investigation agents proposed in production environments across many vendors, each accepted or rejected by a security expert who then wrote down the rules behind those decisions. Each challenge becomes one question with its own measurement. For effectiveness, we ask to what extent the patterns learned in single organizations can be used in the shared model. We compare the model with and without the batch's link types on the asset types they relate, the paths they shorten and the attack paths they extend. For precision, we ask how many of the agents' proposals are correct and whether a correct one can be recognized. We measure the share the expert accepts and the edits it needed, and test whether acceptance can be predicted from the structure of a candidate's query, such as the fields it matches two asset types on, or from how often and in how many environments it was proposed. For scale, we ask how much of the review the rules can take over. We replay the batch with the rules applied before review and measure how many candidates the automatic checks would remove, how often the language-model judge agrees with the expert, and how often a repair step reproduces the expert's edits, each against the accepted link types it would lose.

The rest of the paper is structured as follows. Section~2 places the flywheel among related work, Section~3 describes how the brain structures its knowledge and introduces a running example, and Section~4 defines the flywheel step by step. Section~5 evaluates it, Section~6 discusses what the results mean and future work, and Section~7 concludes.

\section{Related Work}
\label{sec:related}

Systems whose knowledge grows with use differ in five properties: what grows, where it comes from, who shares it, what admits it, and whether admission gets cheaper with use. We use these properties to position the context flywheel, first against such loops in general and then against the five roles they play in security, the flywheel's setting: detection and triage, threat intelligence, detection engineering, attack and environment modeling, and agentic security operations.

Outside security, growth with use is common, but in the loops we review, what grows is a system's own models, skills, indexes or joins, kept within one deployment, agent or database. Data flywheels retrain a deployed system's models on feedback from its own use \citep{zhao2025aitl}; in one, expert analysis of failures became a bottleneck \citep{shukla2026adaptive}, the cost that the flywheel's inner loop targets. An agent keeps the skills that pass its own verification \citep{wang2024voyager}, and an agent ecosystem shares a skill across its agents once a language model approves it \citep{ma2026skillclaw}. Data systems derive indexes from the queries each database runs \citep{das2019automatically} and recover from query history \citep{tripathi2026,chen2025text} the joins that schema documentation omits \citep{shkapenyuk2025}. Such feedback loops \citep{sculley2015hidden} also need anchors against their own errors: models trained on generated data degrade less when original data is kept \citep{shumailov2024ai}, and word lists grown from text drift less when new words are checked against trusted seed words and early additions \citep{mcintosh2009reducing}. In the flywheel, expert review is that anchor.

In security, detection and triage make review cheaper in two ways that the flywheel also uses: they reuse analysts' decisions, and they settle cases automatically before review. Detectors retrain on analysts' labels \citep{veeramachaneni2016ai2}. A triage system learns from analysts' actions to close benign alerts, and in a managed security service, one operations center serving many organizations, it can learn across the organizations it monitors \citep{turcotte2025aact}. New events that match a cluster inherit the score that operators gave it once \citep{vanede2022deepcase}. Review can also be cheap from the start: an automatic first check settles most phishing reports from employees, leaving administrators few decisions \citep{lain2022phishing}. In these systems, what grows is labels and classifiers within one operations center, not a model of the environment.

Threat intelligence shows both the value of sharing across organizations and why agreement between contributors is weaker evidence than their number suggests. Organizations share indicators of compromise \citep{wagner2016misp} and abuse reports \citep{thomas2016abuse}. Open scores of how likely each vulnerability is to be exploited \citep{jacobs2021exploit} are computed, in a later version, by a model trained on exploitation evidence that data-sharing organizations pool \citep{jacobs2023enhancing}. Shared knowledge needs admission: shared feeds vary in coverage and accuracy \citep{li2019reading}. Contributions can also be poisoned: fake threat-intelligence reports pass as real with threat hunters \citep{ranade2021generating}, and compromised participants corrupt a jointly trained detector \citep{nguyen2020poisoning}. Agreement can mislead: in a community that confirms phishing reports by vote, a few members cast most votes \citep{moore2008evaluating}, and anti-malware engines that label alike mislead when counted as independent votes \citep{zhu2020measuring}. Beyond security, copying can carry false values into the majority \citep{dong2009integrating}, and accurate language models err alike even across providers \citep{kim2025correlated}. The number of investigations that propose a candidate therefore says less about whether it is correct than it seems; in the batch we evaluate, recurrence did not predict the expert's decision.

Detection engineering grows structure that serves many organizations, as the flywheel's link types do, and so does attack and environment modeling. The systems we review in both grow that structure from published knowledge rather than from investigations. Detection rules are generated from threat intelligence and validated by a language-model judge \citep{ghaffarzadegan2026}. Attack graphs \citep{sheyner2002} and logical attack models \citep{ou2005mulval} need a rule for each attack technique, written by experts or derived from vulnerability descriptions \citep{binyamini2021framework}. Security knowledge graphs built from threat reports and public vulnerability databases describe threats in general, not an organization's assets \citep{cheng2025ctinexus,xu2026trace}. The brain's own link types are also built before investigations need them \citep{engelberg2026asset}.

Agentic security operations come closest in where knowledge comes from: agents keep or reuse what investigations find. They hold knowledge-graph memory of the hosts, ports, services and vulnerabilities found during a penetration test \citep{wang2026ptfusion}, fuse past alerts and tickets with live alerts \citep{wang2026defengraph}, or propose updates to detection rules and response playbooks that an audit admits \citep{mitra2026agenticcyops}. That memory stays within one organization or deployment, and we found no report that admitting it gets cheaper.

Together, these lines of work leave one combination unaddressed: a structural model of the environment that grows from the relationships investigations discover, is shared across organizations as type-level patterns, and is admitted by experts whose decisions become automatic checks. The context flywheel is built for this combination, to close the \emph{modeling gap}. Applying expert judgment automatically has precedents outside security: labeling functions, heuristics in which experts encode their judgment to label data at scale \citep{ratner2017snorkel}, and strong language-model judges, which agree with humans about as often as humans agree among themselves \citep{zheng2023judge}. Reviewers refine their criteria as they grade \citep{shankar2024}; in the flywheel, the expert writes the rules after each batch, from the criteria applied in it. In this paper, the reviews of one batch become checks replayed on that same batch; whether admission gets cheaper from batch to batch remains to be measured.

\section{Knowledge Structure in the Enterprise Security Brain}

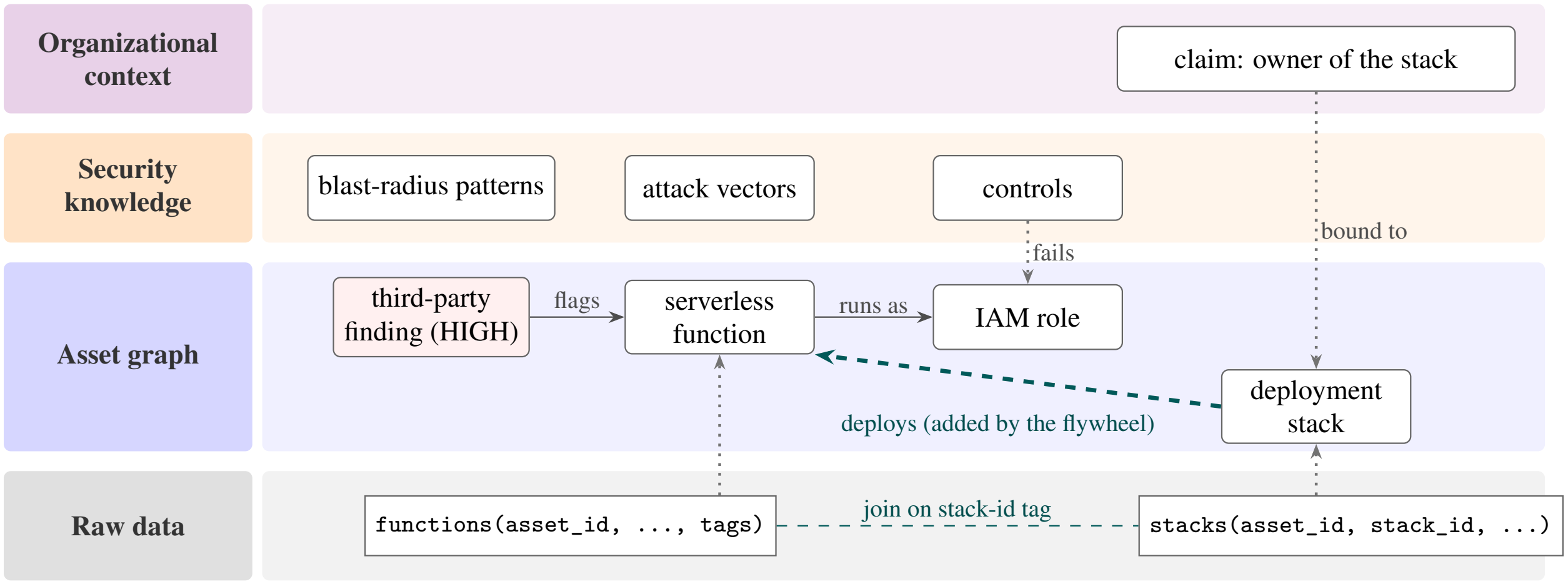
\begin{figure}[H]
\centering
\resizebox{\linewidth}{!}{%
\begin{tikzpicture}[
  font=\footnotesize,
  band/.style={rounded corners=2pt, minimum width=12.9cm, inner sep=0pt},
  lbl/.style={rounded corners=2pt, minimum width=2.5cm, align=center, font=\footnotesize\bfseries, text=black!80, inner sep=2pt},
  n/.style={draw=black!60, rounded corners=2pt, align=center, inner sep=3pt, fill=white, minimum height=6.5mm, minimum width=19mm},
  tb/.style={draw=black!60, font=\scriptsize\ttfamily, inner sep=3pt, fill=white, minimum height=6mm},
  e/.style={-{Stealth[length=1.6mm]}, black!70},
  el/.style={font=\scriptsize, inner sep=1pt, text=black!70},
  xa/.style={dotted, thick, black!55, -{Stealth[length=1.5mm]}},
  newl/.style={very thick, dashed, teal!70!black, -{Stealth[length=2mm]}}]
\def\bx{9.05}
\node[band, fill=violet!7,  minimum height=1.1cm] (B4) at (\bx,6.05) {};
\node[band, fill=orange!8,  minimum height=1.1cm] (B3) at (\bx,4.75) {};
\node[band, fill=blue!6,    minimum height=1.9cm] (B2) at (\bx,3.05) {};
\node[band, fill=black!5,   minimum height=1.1cm] (B1) at (\bx,1.35) {};
\node[lbl, fill=violet!18, minimum height=1.1cm] at (1.25,6.05) {Organizational\\context};
\node[lbl, fill=orange!22, minimum height=1.1cm] at (1.25,4.75) {Security\\knowledge};
\node[lbl, fill=blue!16,   minimum height=1.9cm] at (1.25,3.05) {Asset graph};
\node[lbl, fill=black!12,  minimum height=1.1cm] at (1.25,1.35) {Raw data};
\node[n, minimum width=40mm] (O) at (13.2,6.05) {claim: owner of the stack};
\node[n] (BR) at (4.3,4.75) {blast-radius patterns};
\node[n] (AV) at (7.2,4.75) {attack vectors};
\node[n] (CT) at (10.3,4.75) {controls};
\node[n, fill=red!6] (X) at (4.3,3.45) {third-party\\finding (HIGH)};
\node[n] (F) at (7.2,3.45) {serverless\\function};
\node[n] (R) at (10.3,3.45) {IAM role};
\node[n] (S) at (13.2,2.55) {deployment\\stack};
\draw[e] (X) -- node[el, above]{flags} (F);
\draw[e] (F) -- node[el, above]{runs as} (R);
\draw[newl] (S.west) -- (F.south east);
\node[el, text=teal!60!black] at (10.0,2.36) {deploys (added by the flywheel)};
\draw[xa] (CT) -- node[el, right]{fails} (R);
\draw[xa] (O) -- node[el, right]{bound to} (S);
\node[tb] (TF) at (5.7,1.35) {functions(asset\_id, \ldots, tags)};
\node[tb] (TS) at (13.55,1.35) {stacks(asset\_id, stack\_id, \ldots)};
\draw[dashed, teal!70!black] (TF.east) -- node[el, above, text=teal!60!black]{join on stack-id tag} (TS.west);
\draw[xa] (TF.north -| F) -- (F.south);
\draw[xa] (TS.north -| S) -- (S.south);
\end{tikzpicture}}
\caption{The four tiers of the brain, shown at the instance level for one organization in the running example; dashed: the link type the flywheel adds and the rule it reads.}
\label{fig:tiers}
\end{figure}

The brain holds knowledge in four tiers, each derived from the tier beneath it \citep{yavin2026context} (Figure~\ref{fig:tiers}). The raw data tier holds what connected products report, their configurations, metadata and findings, as versioned, vendor-neutral tables \citep{engelberg2026asset,goldberg2026coding}. A \emph{finding} is a product's report that a resource breaks one of its rules, such as a scanner flagging a function that any caller can invoke; a \emph{third-party finding} comes from a product other than the brain.

The asset graph holds knowledge at two levels. The \emph{type level} is the model, the same for every organization. Its nodes are asset types, such as a serverless function (code that a cloud provider runs on demand), an IAM role (an identity that workloads assume to obtain permissions) or a deployment stack (the cloud resources created together from one template). Its edges are \emph{link types}: relationship patterns between two asset types, written as rules over the raw data, such as ``a role can assume a role'' or ``a stack deploys a function''. We call this graph the \emph{type-level graph}. The \emph{instance level} is the model's manifestation over one organization's data: each resource becomes an asset of its type, each pair of assets that satisfies a link type's rule becomes a \emph{link instance}, and findings attach to the assets they concern \citep{engelberg2026asset,goldberg2026coding}.

Security knowledge, the third tier, is defined over types and evaluated over instances, as executable specifications run against an organization's asset graph when a question arrives \citep{engelberg2026asset,goldberg2026coding}. It takes three forms: a \emph{control} marks an asset as failing a security requirement, such as a role that grants administrative access; an \emph{attack vector} is a path along link types that an attacker can follow, such as from a role to every role it can assume; and a \emph{blast-radius pattern} collects what a compromised asset can reach. The fourth tier, organizational context, is instance level: claims about ownership, business function and accepted risk, bound, where a binding exists, to one organization's assets \citep{yavin2026context}. Because the model is defined at the type level, a link type admitted once yields link instances in every organization whose data holds both asset types, and every control, attack vector and blast-radius pattern that crosses those types is evaluated over them.

Our running example is the contextualization of a third-party finding \citep{goldberg2026findings}. A finding $f$ flags a resource $r$ with a severity $s_v$ that the vendor's rule fixed before it met any environment; contextualization returns an adjusted severity $s$ for the environment where $r$ lives, with an evidence trace $E$ of the facts the decision rests on. Evidence comes in three kinds \citep{goldberg2026findings}: the resource's own state $\sigma(r)$, its neighborhood $N_k(r)$ of assets within $k$ hops in the asset graph, and the signals $X(r)$ that other products hold about $r$. In production, the deciding evidence for about half of the findings lay outside the flagged resource \citep{goldberg2026findings}. The agent that performs the task is a language model with four actions on the brain: a \emph{structured query} over the raw data tables, which can join across products; a \emph{graph query}, which returns $N_k(r)$ with the controls that neighboring assets fail; a \emph{neighborhood exploration}, which lists the asset types that an asset type links to; and a \emph{catalog search}, which finds data other products hold about $r$, including data the graph does not link.

\begin{figure}[tbp]
\centering
\resizebox{\linewidth}{!}{%
\begin{tikzpicture}[
  font=\footnotesize,
  q/.style={draw=black!55, rounded corners=2pt, align=center, fill=white, text width=27mm, minimum height=25mm, inner sep=2pt, execute at begin node={\hyphenpenalty=10000\exhyphenpenalty=10000}},
  qb/.style={q, draw=blue!50!black},
  qa/.style={q, draw=teal!70!black, thick},
  term/.style={draw=black!60, rounded corners=6pt, align=center, fill=black!4, minimum height=25mm, text width=27mm, inner sep=2pt, execute at begin node={\hyphenpenalty=10000\exhyphenpenalty=10000}},
  e/.style={-{Stealth[length=1.6mm]}, black!65},
  ce/.style={-{Stealth[length=1.6mm]}, dashed, teal!70!black},
  lane/.style={font=\scriptsize\bfseries, inner sep=2pt}]
\node[term, fill=red!6] (F) at (-0.8,0) {finding $f$: any caller can invoke function $r$ (HIGH)};
\node[qb] (Q1) at (2.5,0) {\textbf{Q1} structured query\\{\scriptsize\itshape read who may invoke the function}\\{\scriptsize Returns: any caller}};
\node[qb] (Q2) at (5.65,0) {\textbf{Q2} graph query\\{\scriptsize\itshape find the role it runs as and the controls that role fails}\\{\scriptsize Returns: a role that reads one queue; no failed control}};
\node[qb] (Q3) at (8.8,0) {\textbf{Q3} graph query\\{\scriptsize\itshape list what that role can reach}\\{\scriptsize Returns: one queue; no sensitive data}};
\node[qb] (Q4) at (11.95,0) {\textbf{Q4} neighborhood exploration\\{\scriptsize\itshape list the asset types a function links to}\\{\scriptsize\color{red!70!black} Returns: no deployment stack}};
\node[qb] (Q5) at (15.1,0) {\textbf{Q5} catalog search\\{\scriptsize\itshape find where the stack that created a resource is recorded}\\{\scriptsize Returns: the stacks table and a stack-id tag on functions}};
\node[qa] (Q6) at (15.1,-3.6) {\textbf{Q6} agent query\\{\scriptsize\itshape match each function to the stack named in its stack-id tag}\\{\scriptsize Returns: a sandbox stack}};
\node[qa] (Q7) at (11.95,-3.6) {\textbf{Q7} agent query\\{\scriptsize\itshape match each function to the distributions in its account}\\{\scriptsize Returns: every distribution in the account}};
\node[term] (V) at (8.8,-3.6) {severity lowered to MEDIUM; evidence $E$ from Q1, Q2, Q3 and Q6};
\node[term, fill=teal!8, draw=teal!70!black, text width=62mm, minimum height=10mm] (C) at (13.5,-6.4) {candidate link types, collected by the flywheel};
\draw[e] (F) -- (Q1); \draw[e] (Q1) -- (Q2); \draw[e] (Q2) -- (Q3); \draw[e] (Q3) -- (Q4); \draw[e] (Q4) -- (Q5);
\draw[e] (Q5) -- (Q6); \draw[e] (Q6) -- (Q7); \draw[e] (Q7) -- (V);
\draw[ce] (Q6.south) -- node[right, font=\scriptsize, text=teal!60!black]{kept} (Q6.south |- C.north);
\draw[ce] (Q7.south) -- node[left, font=\scriptsize, text=red!60!black]{dropped} (Q7.south |- C.north);
\begin{scope}[on background layer]
\node[fill=blue!5, rounded corners=3pt, fit=(Q1)(Q2)(Q3)(Q4)(Q5), inner sep=5pt] (L1) {};
\node[fill=teal!6, rounded corners=3pt, fit=(Q6)(Q7), inner sep=5pt] (L2) {};
\end{scope}
\node[lane, text=blue!50!black, anchor=south west] at (L1.north west) {retrieved from the brain's knowledge};
\node[lane, text=teal!60!black, anchor=south west] at (L2.north west) {written by the agent};
\end{tikzpicture}}
\caption{Investigation graph of the running example, with each query described in words (simplified).}
\label{fig:investigation}
\end{figure}
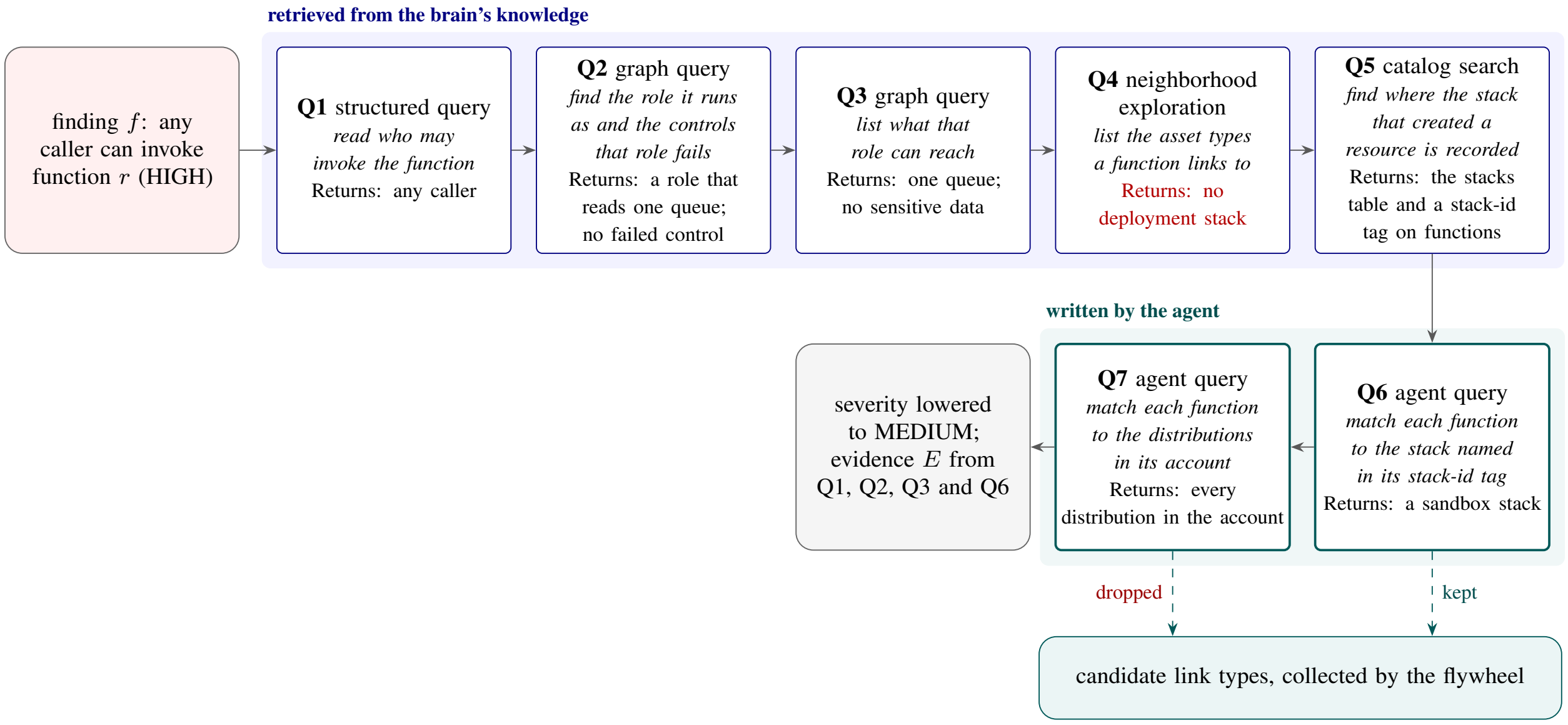

In the example, a cloud security platform flags a serverless function $r$ as invocable by any caller, with severity HIGH, and the agent must decide whether that exposure is dangerous in this environment. Figure~\ref{fig:investigation} shows the investigation as a graph of queries, each described in words together with what it returns. The example is a composite: the link type it leads to and that link type's review come from the evaluated batch, while the finding, the other queries and their results are simplified and illustrative.

The agent first retrieves from the brain's knowledge. Q1, a structured query, confirms the finding: the function's policy lets any caller invoke it. Q2, a graph query, asks what an attacker would gain by invoking it: the function runs as a role that reads one message queue, and the role fails no control. Q3 evaluates the blast-radius pattern of that role and finds no sensitive data store within reach. One question remains, whether the function belongs to a production application, and the deployment stack that created it would answer it. Q4 explores what functions link to and finds no link type to deployment stacks. Q5, a catalog search, finds the stacks' table and the tag in which the deployment service records, on each resource it creates, the stack that created it. The agent then writes its own queries. Q6 joins functions to stacks on that tag and returns a sandbox stack. Q7, written to find what serves the function, joins functions to content-delivery distributions on the account they share and returns every distribution in the account, which answers nothing. With limited reach and a sandbox deployment, the agent lowers the severity to MEDIUM and records Q1, Q2, Q3 and Q6 as evidence.

Agent-written queries such as Q6 and Q7 are what the flywheel collects. Without it, both are discarded when the investigation ends, and the next investigation of a finding on a function, in any organization, writes them again. They also differ in kind: Q6 records a relationship that holds for every function a stack creates, and Q7 records only that two assets share an account. The flywheel has to keep the first and drop the second. Section~4 follows both through the flywheel.

\section{The Context Flywheel}

\begin{figure}[H]
\centering
\begin{tikzpicture}[
  box/.style={draw, rounded corners=2pt, align=center, minimum height=12mm, minimum width=25mm, font=\small},
  main/.style={-{Stealth[length=2mm]}, thick},
  inner/.style={-{Stealth[length=2mm]}, dashed},
  lab/.style={font=\scriptsize, align=center},
  st/.style={font=\scriptsize\itshape, text=gray!60!black}]
\node[lab] (Q) at (-2.5,0) {questions\\and findings};
\node[box] (A) at (0,0) {Investigation\\agents\\[-1pt]{\scriptsize\itshape 1 proposal}};
\node[box] (K) at (3.8,0) {Deterministic\\checks\\[-1pt]{\scriptsize\itshape 2 checks}};
\node[box] (R) at (7.6,0) {Expert\\review\\[-1pt]{\scriptsize\itshape 3 review}};
\node[box, fill=teal!10] (M) at (11.4,0) {Security brain\\(shared model)\\[-1pt]{\scriptsize\itshape 4 propagation}};
\node[box] (W) at (3.8,-2.7) {Written rules\\[-1pt]{\scriptsize\itshape 5 learning from review}};
\draw[main] (Q) -- (A);
\draw[main] (A) -- node[lab, above]{candidate\\link types} (K);
\draw[main] (K) -- (R);
\draw[main] (R) -- node[lab, above]{accepted} (M);
\draw[main] (M.north) -- ++(0,0.7) -| node[lab, above, pos=0.25]{richer model, read by every later investigation} (A.north);
\draw[inner] (R.south) |- node[lab, right, pos=0.35]{written down\\after each batch} (W.east);
\draw[inner] (W.north east) -- node[lab, above, sloped]{judge, repair} (R.south west);
\draw[inner] (W.north) -- node[lab, left]{rule checks} (K.south);
\draw[inner] (W.west) -| node[lab, below, pos=0.25]{instructions} (A.south);
\end{tikzpicture}
\caption{The five steps of the context flywheel: the funnel (solid) and the inner loop (dashed).}
\label{fig:flywheel}
\end{figure}
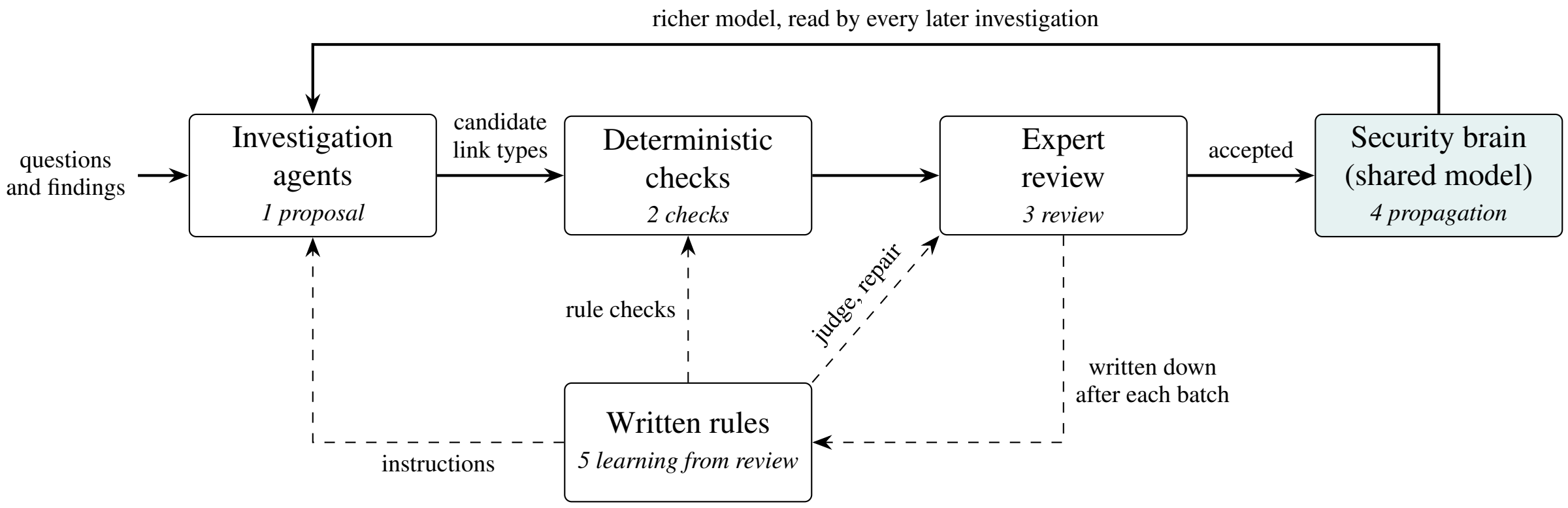

The context flywheel increases the brain's knowledge with use: it turns the relationships that agents work out into link types of the shared model, and it turns each review of those candidates into rules that take over part of the next review. It runs in five steps (Figure~\ref{fig:flywheel}). The first three form a funnel that many candidates enter and few leave, and every step is automatic except the expert's decision. A query can add a link type or a control to the brain. This paper studies link types, since missing relations account for most of the tasks on which the brain did not lead \citep{goldberg2026coding}. We follow Q6 and Q7 of the running example through the steps.

In step~1, proposal, an agent that connects two asset types the brain does not link submits the connection as a candidate link type, with its explanation, while the investigation runs. The funnel records how often and in how many environments each candidate is proposed. In the running example, the agent submits Q6 as ``stack deploys function'' and Q7 as a link type between functions and content-delivery distributions.

In step~2, automatic checks remove candidates that can never be admitted: those tied to one organization, so that no organization's data enters the shared model; those that are malformed or name unknown asset types; those that return nothing; and those the brain already holds. The checks test whether a candidate can be admitted, not whether it is good, so both Q6 and Q7 pass.

In step~3, a security \emph{expert} accepts, edits or rejects each remaining candidate, judging what the checks cannot: whether the connection can join an asset to the wrong asset, and whether it records a relationship or a coincidence. Review favors precision, since a wrong link type misleads every later investigation in every organization. In the running example, the expert accepts Q6, after correcting how it reads the stack identifier: each function carries the identifier of the stack that created it, so matching on that identifier connects every function to exactly one stack, the right one. The expert rejects Q7: two assets in the same account need not be related, so matching on the account connects every function to every content-delivery distribution in that account, whether or not the distribution serves the function.

In step~4, propagation, the accepted candidate becomes a link type of the shared model, and every organization whose data holds both asset types gains the matching link instances. Security knowledge, organizational context and agents read them without further work: a blast-radius pattern that starts from a compromised stack now reaches its functions, and the next investigation of a function finds its stack with one graph query. Each admitted link type turns evidence that investigations had to work out into evidence they can read, so later candidates point at the gaps that remain. We describe this compounding; the evaluation measures the model's structure, not later investigations.

In step~5, learning from review, the expert writes down the criteria applied in a batch as short rules, and the flywheel reuses them in three places. They become instructions to the proposing agents; \emph{rule checks} that either reject a candidate before review (\emph{clear-cut} checks) or warn the expert (\emph{softer} checks); and the rubric of a language-model judge, which grades candidates before review, and of a repair step, which proposes the expert's edit. A deterministic fix also repairs one frequent defect automatically. In the running example, the rejection of Q7 becomes the rule ``prefer strong keys'', whose check would remove the next candidate that links two asset types only through a shared account before the expert sees it.

\section{Evaluation}

We evaluate the flywheel by replaying a sample of a production review batch through its steps and comparing each outcome with an expert's decisions. The evaluation answers the three challenges of the introduction. The protocol defines the criteria for each challenge and how each is measured, the setup describes the batch and the components evaluated, and the results follow the criteria in the protocol's order.

\subsection{Protocol}

The protocol states each challenge as a question, splits it into criteria, and gives each criterion a measure, illustrated with the running example. We denote by $C$ the candidates in the batch, by $A$ those the expert accepted and by $R = C - A$ those rejected.

\paragraph{Effectiveness.} To what extent can a link type learned in one organization be used in the shared model? The shared model uses a link type in three ways, each a criterion, measured on the type-level graph without the accepted link types, $G_0$, and with them, $G_1$, as an absolute and a relative change. \emph{Coverage} asks whether a link type relates asset types that the model left unrelated. Q6, for example, relates serverless functions to deployment stacks, which no link type joined before (Q4 in Figure~\ref{fig:investigation}). We count the asset types with at least one link type and the pairs of asset types joined directly by one. \emph{Connectivity} asks whether a link type shortens the paths that queries traverse between asset types. We count the asset types in the largest connected component and compute global efficiency \citep{latora2001}, the mean inverse path length over the $n$ asset types:
\[
E(G) = \frac{1}{n(n-1)} \sum_{u \neq v} \frac{1}{d(u,v)},
\]
where $d(u,v)$ is the number of link types on the shortest path between asset types $u$ and $v$ in either direction, and $1/d(u,v) = 0$ when no path exists. $E(G)$ lies between 0 and 1. Because an asset type with no link type adds zero to every term it enters, we also compute $E(G)$ over the asset types linked before the batch, which separates shorter paths from newly linked asset types. A link type that joins two asset types previously four link types apart, for example, raises their term from $1/4$ to 1. \emph{Attack paths} asks whether a link type extends the paths that attack-path analysis can follow. We count the attack-path reach: the ordered pairs $(u, v)$ such that $v$ can be reached from $u$ along link types an attacker can traverse, followed in their direction. A link type through which an identity can assume a role, for example, makes that role, and every role it can assume in turn, reachable from the identity.

\paragraph{Precision.} How many of the agents' proposals are correct, and can a correct one be recognized before review? \emph{Acceptance} asks what share of the proposals the expert admits, and how many need correcting first. The expert accepts Q6 after correcting how it reads the stack identifier and rejects Q7. We report $|A|/|C|$ with a 95\% Wilson interval \citep{wilson1927}, counting a candidate as accepted if the model gained a link type between the same two asset types. We also report the share of $A$ the expert edited, and the share of those edits reproduced exactly by one automatic fix, which makes every optional join require both ends. \emph{Predictability} asks whether the expert's decision can be told before review, and from which signal. Three signals exist before review. \emph{Recurrence} is how often and in how many environments a candidate was proposed. \emph{Vendor} is the vendor whose asset types the candidate links. \emph{Query structure} is how the candidate's query relates the two asset types: the fields it matches them on, such as a unique identifier or only a shared account, and the tables it passes through, such as a log of events; the rule checks read it from the query without running it. Recurrence matters most: if link types that many environments propose were usually correct, they could be admitted with less review. Yet a query can recur and still be wrong. Q7 could be proposed in every environment where functions and distributions share an account, and only its structure, a match on the account alone, marks it as wrong; Q6 matches each function to the one stack its tag names. For each rule check $k$, with $F_k$ the candidates it flags, we report the acceptance rate $|A \cap F_k|/|F_k|$, which is near 0 when the check marks rejections reliably. For each signal, and for all together, a simple classifier is trained on four fifths of the candidates to predict the decision and scored on the remaining fifth, over 20 splits. The score is $\mathrm{AUC} = \Pr[s(a) > s(r)]$, the probability that the classifier ranks a held-out accepted candidate $a$ above a held-out rejected candidate $r$: 0.5 means the signal tells nothing about the decision, and 1 means it separates the two perfectly.

\paragraph{Scale.} How much of the expert's review can the rules take over? \emph{Filtering rate} asks how much of the queue the rule checks can remove before review, and how many good link types go with it. The check derived from ``prefer strong keys'', for example, removes Q7 before the expert sees it; a check that also removed Q6 would lose a good link type. Only the clear-cut checks, those used to reject, act as the gate. With $F$ the candidates they flag, we report the share of the queue removed, $|F|/|C|$, the share of accepted link types lost, $|F \cap A|/|A|$, and the acceptance rate among the candidates left, $|A - F|/|C - F|$. Because the clear-cut group was chosen after seeing the batch, it is also chosen on four fifths of the candidates and evaluated on the rest, over 100 splits. \emph{Agreement rate} asks how often a language-model judge given the rules grades candidates as the expert decided them. A judge that rejects Q7 and accepts Q6 after correction agrees with the expert on both. We report it corrected for chance, as Cohen's $\kappa = (p_o - p_e)/(1 - p_e)$ \citep{cohen1960}, where $p_o$ is the raw agreement rate, the share of candidates on which the judge and the expert agree, and $p_e$ the rate expected by chance; 0 is chance agreement and 1 perfect agreement. \emph{Repair} asks whether a repair step given the rules makes the expert's edits. For Q6, it succeeds if it corrects how Q6 reads the stack identifier as the expert did. We report the share of $A$ for which the repair step's output equals the expert's accepted query, exactly or, where the two differ in form, as judged by both judge models.

The effectiveness measures show what patterns learned in single organizations add to the shared model, not in how many organizations each yields link instances. The rules were written after this review, from these decisions, so every result that uses them measures how well the expert's criteria account for this batch, not how well they predict the next one. Choosing the clear-cut group on held-out splits removes the effect of selecting that group from the data, not the effect of writing the rules after the review.

\subsection{Setup}

We replay a sample of a production review batch, because the flywheel's input is the queries that real investigations write and its ground truth is an expert's decision on each of them. The sample, which we call the batch, holds the 656 candidate link types that investigation agents proposed in 21 production environments and that passed the automatic checks; they cover 15 vendors, and 447 of them (68\%) are AWS link types. A candidate is one proposed link type with its query and the agent's explanation, and repeated proposals count as one candidate with its recurrence. A security expert reviewed the whole batch and recorded every accepted link type with its edits; a candidate not accepted counts as rejected.

The judge combines two language models from one provider, Claude Opus 5.5 and Claude Sonnet 5.5, each run once on the query, the agent's explanation, the columns of the tables involved and the rules, without seeing the expert's decision. Each grades a candidate as reject, accept after correction or accept as is, and the combined grade is the lower of the two, a strict combination that suits a gate favoring precision. The repair step runs Claude Opus 5.5 under the same rules and keeps, rewrites or rejects each query.
\subsection{Results}

\begin{figure}[H]
\centering
\tikzset{
  n/.style={draw, rounded corners=2pt, font=\footnotesize, minimum height=6mm, inner sep=3pt},
  endn/.style={n, thick, fill=teal!10},
  old/.style={dashed, gray},
  el/.style={font=\scriptsize, text=gray!60!black, below},
  newl/.style={very thick, teal!70!black, -{Stealth[length=2mm]}},
  nl/.style={font=\scriptsize\bfseries, text=teal!60!black, above}}
\begin{subfigure}{\linewidth}
\centering
\begin{tikzpicture}
\node[endn] (a1) at (0,0) {LB listener};
\node[n] (a2) at (3.5,0) {third-party finding};
\node[n] (a3) at (7.1,0) {AWS access key};
\node[n] (a4) at (10.3,0) {Okta user};
\node[endn] (a5) at (13.3,0) {Okta application};
\draw[old] (a1) -- node[el]{has finding} (a2);
\draw[old] (a2) -- node[el]{has finding} (a3);
\draw[old] (a3) -- node[el]{matched to} (a4);
\draw[old] (a4) -- node[el]{can open} (a5);
\draw[newl] (a1.north) to[bend left=14] node[nl]{new link type: authenticates through} (a5.north);
\end{tikzpicture}
\caption{Load balancer listener to the Okta application it authenticates through.}
\label{fig:shortcut-okta}
\end{subfigure}

\vspace{4mm}
\begin{subfigure}{\linewidth}
\centering
\begin{tikzpicture}
\node[endn] (c1) at (0,0) {compute instance};
\node[n] (c2) at (3.9,0) {AWS account};
\node[n] (c3) at (7.7,0) {Kubernetes cluster};
\node[endn] (c4) at (11.4,0) {node group};
\draw[old] (c1) -- node[el]{in account} (c2);
\draw[old] (c2) -- node[el]{has cluster} (c3);
\draw[old] (c3) -- node[el]{has group} (c4);
\draw[newl] (c1.north) to[bend left=14] node[nl]{new link type: is node of} (c4.north);
\end{tikzpicture}
\caption{Compute instance to the Kubernetes node group it runs in.}
\label{fig:shortcut-eks}
\end{subfigure}
\caption{Paths before the batch (dashed) and the link types the batch added (solid).}
\label{fig:shortcuts}
\end{figure}
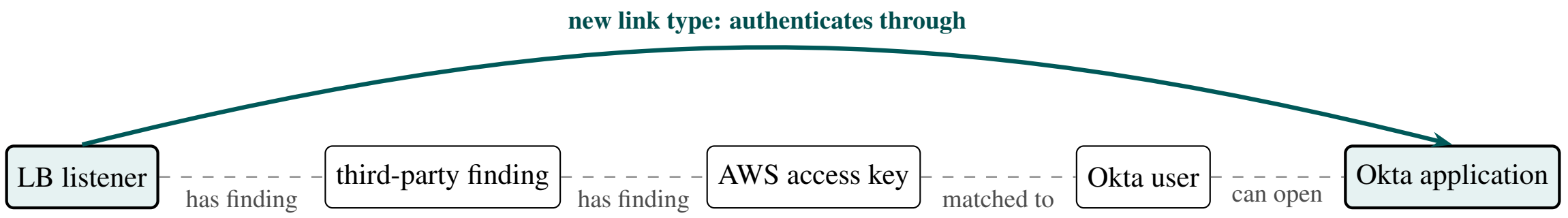
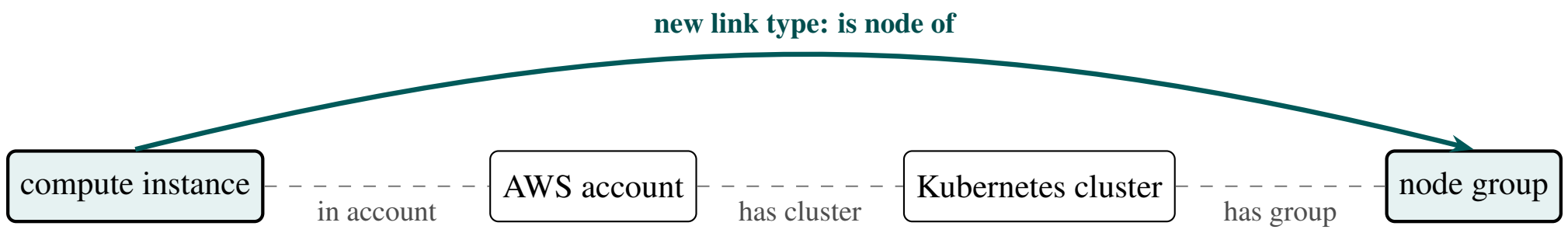

\paragraph{Effectiveness.} The accepted link types extend the shared model on all three criteria (Table~\ref{tab:graph}). For \emph{coverage}, the batch brings 49 asset types from 7 vendors into the type-level graph; before it, no link type reached them. The pairs of asset types joined directly grow by 9\%, most for the asset types the batch touched, whose median number of directly linked asset types rises from 3 to 5. For \emph{connectivity}, the largest connected component grows from 765 to 816 asset types and global efficiency rises by 13\%, mostly because the 49 asset types that coverage adds now have paths to the rest of the graph. Among the 786 asset types linked before the batch, efficiency rises by 2\%, from 0.289 to 0.295: 72 link types join asset types that had no path between them, and 43 replace a path of three or more link types. Before the batch, for example, load balancer listeners reached the Okta applications they authenticate through only by four link types that connect them by coincidence, and compute instances reached their Kubernetes node groups only through a shared account; one link type each now joins them directly (Figure~\ref{fig:shortcuts}). For \emph{attack paths}, attack-path reach grows by 19\%, from 26,934 to 32,030 ordered pairs of asset types; the link type from a GitHub organization to the cloud roles it can assume, for example, extends attack-path analysis to start from GitHub.

\begin{table}[H]
\centering
\small
\caption{The type-level graph (835 asset types) without and with the accepted link types.}
\label{tab:graph}
\begin{tabular}{@{}llrrr@{}}
\toprule
Criterion & Measure & Without & With & Change \\
\midrule
Coverage & Asset types with at least one link type & 786 & 835 & +49 (6.2\%) \\
 & Directly linked pairs of asset types & 2,660 & 2,900 & +240 (9.0\%) \\
Connectivity & Asset types in the largest connected component & 765 & 816 & +51 (6.7\%) \\
 & Global efficiency $E(G)$ & 0.256 & 0.290 & +0.034 (13.3\%) \\
 & Global efficiency among the 786 asset types linked before & 0.289 & 0.295 & +0.006 (2.1\%) \\
Attack paths & Attack-path reach (ordered pairs of asset types) & 26,934 & 32,030 & +5,096 (18.9\%) \\
\bottomrule
\end{tabular}
\end{table}

\begin{figure}[H]
\centering
\includegraphics[width=\linewidth]{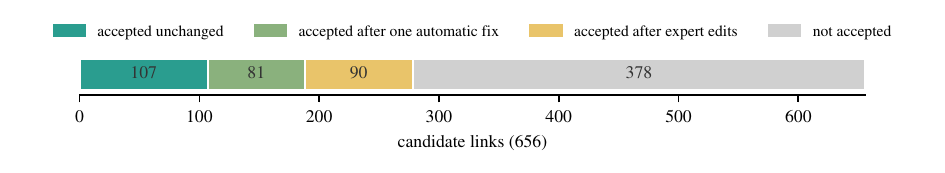}
\caption{Expert review of the 656 candidate link types.}
\label{fig:outcome}
\end{figure}

\paragraph{Precision.} The expert accepts fewer than half of the agents' proposals, and only query structure predicts which. For \emph{acceptance}, the expert accepted 278 of the 656 candidates (42\%, 95\% interval [39\%, 46\%]) and edited 171 of them; the automatic fix reproduces 81 of those edits, so 68\% of accepted link types need no hand edit, and the other 90 edits mostly change how the two asset types are matched (Figure~\ref{fig:outcome}). For \emph{predictability}, a classifier scored on held-out candidates reaches an AUC of 0.78 with the rule checks, against 0.51 for recurrence, which is chance, and 0.59 for vendor; all signals together reach 0.79. Recurrence says nothing about correctness: a link type proposed five or more times is accepted about as often as one proposed once (42\% and 47\%). One untested explanation is that a flawed query written for a common finding recurs wherever that finding occurs. Vendor predicts little, since two thirds of the candidates come from one cloud provider whose acceptance rate is close to the overall one. Query structure predicts through the clear-cut checks, which flag 44\% of rejected candidates and 3\% of accepted ones; the most common, a match only on a shared account, region or project, as in Q7, accounts for a third of all rejections. The softer checks flag accepted and rejected candidates alike, and one in five rejections breaks no check, so part of the expert's judgment cannot be read from the query. The rule-check result is optimistic, since the rules were written from these decisions.

\begin{figure}[H]
\centering
\includegraphics[width=\linewidth]{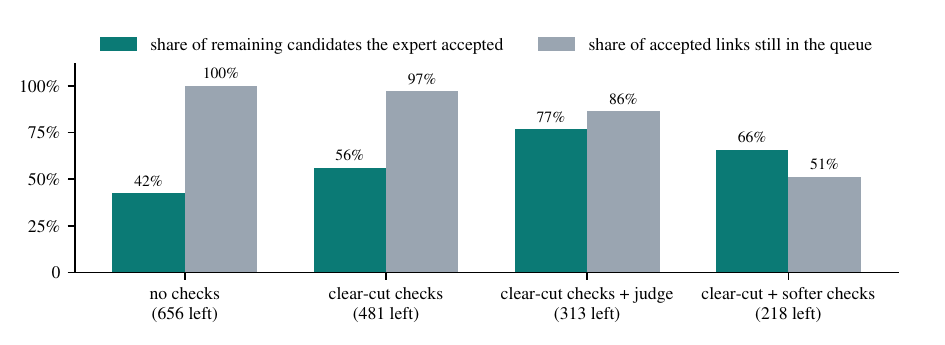}
\caption{Checks before review: acceptance among remaining candidates and accepted link types retained.}
\label{fig:gates}
\end{figure}

\paragraph{Scale.} The rules take over part of the review: the checks remove a quarter of the queue at little loss and the judge agrees with the expert well beyond chance, while the repair step remains a suggestion. For the \emph{filtering rate}, the clear-cut checks run before review would have removed 175 of the 656 candidates (27\%), of which 8 were later accepted (3\% of accepted link types), and the expert would have accepted 56\% of the 481 left instead of 42\% (Figure~\ref{fig:gates}). Chosen on held-out splits, the clear-cut group gives nearly the same result: 25\% of the queue removed for 2.5\% of accepted link types. Adding the softer checks to the gate would remove 49\% of the accepted link types, so they serve better as warnings. For the \emph{agreement rate}, the judge agrees with the expert at $\kappa = 0.62$, against 0.38 for the clear-cut checks alone, and is stricter: of the link types the expert accepted unchanged, it accepts 20\% as they are. Running the clear-cut checks and then the judge would leave 48\% of the candidates for review, of which the expert accepted 77\%, at a cost of 14\% of accepted link types. For \emph{repair}, the repair step reproduces the expert's edit for 28 of the 171 link types the expert edited (16\%). Its 26\% agreement over all 278 accepted link types also counts 44 that the expert accepted unchanged. Of the 107 accepted unchanged, it rewrites 64 (60\%) and rejects 12 (11\%), so it can suggest edits to the expert but not replace them.

\section{Discussion}

The results support three points, one per challenge. First, investigations are a source of model structure. A relationship that an agent works out to settle one finding in one organization, once admitted as a link type, relates asset types the shared model left isolated, shortens some paths between asset types and extends attack-path analysis, for every organization at once. The model therefore grows where investigations need it, and it grows by sharing patterns at the type level, never data. Second, the correctness of a relationship pattern lies in how it relates assets, not in how often agents propose it. If agents that face the same finding write the same flawed query, their agreement is not independent evidence, as with dependent sources of shared threat intelligence (Section~\ref{sec:related}). Admission should therefore be ordered by query structure, and much of the judgment it needs can be written as rules about which fields identify an asset, while the rest stays with the expert. Third, review is itself a source of knowledge. Written as rules, the expert's decisions become checks, instructions and a judge's rubric that can take over the recurring part of the next review, so the expert's attention moves to the decisions no rule yet explains. Together, each batch extends the shared model and moves part of the expert's judgment into the flywheel itself.

Future work follows from what the flywheel is designed to do beyond one batch. Applying the rules to the next batch will measure whether admission gets cheaper from batch to batch, the property that distinguishes the flywheel from loops whose admission cost stays fixed. Counting the link instances that each accepted link type yields in each organization, and the later investigations that use it, will measure how far patterns found in one organization generalize. Extending the flywheel from link types to controls and other security knowledge will test whether the same loop grows every tier of the brain.

\section{Conclusion}

We presented the context flywheel, a system whose knowledge grows with use. It turns the relationships that investigation agents work out and discard into link types of a shared model, through an automatic process with one human step, and it turns each review into rules that take over part of the next one. What it shares across organizations is patterns, never data. In the batch we evaluated, the admitted patterns related asset types the model had left isolated, shortened some paths between asset types and extended attack-path analysis, and the structure of a candidate's query, not its recurrence, predicted the expert's decision. The rules written from one review are applied before the next; on this batch, they would have removed a quarter of the queue at little loss. A security brain built this way grows with its use: each investigation can extend the model that the next one reads.

\bibliographystyle{unsrtnat}
\bibliography{references}

\end{document}